# A mildly relativistic radio jet from the otherwise normal type Ic supernova 2007gr

Z. Paragi[1,2], G. B. Taylor[3], C. Kouveliotou[4], J. Granot[5], E. Ramirez-Ruiz[6], M. Bietenholz[7,8], A. J. van der Horst[4], Y. Pidopryhora[1], H. J. van Langevelde[1,9], M. A. Garrett[9,10,11], A. Szomoru[1], M. K. Argo[12], S. Bourke[1] & B. Paczyński‡

[1]Joint Institute for VLBI in Europe (JIVE), Postbus 2, 7990AA Dwingeloo, The Netherlands. [2]MTA Research Group for Physical Geodesy and Geodynamics, PO Box 91, H-1521 Budapest, Hungary. [3]University of New Mexico, Department of Physics and Astronomy, MSC07 4220, 800 Yale Blvd NE Albuquerque, New Mexico 87131-0001, USA. [4]Space Science Office, NASA/Marshall Space Flight Center, Huntsville, Alabama 38512, USA. [5]Centre for Astrophysics Research, University of Hertfordshire, College Lane, Hatfield, Hertfordshire AL10 9AB, UK. [6]Department of Astronomy and Astrophysics, University of California, Santa Cruz, California 95064, USA. [7]Hartebeesthoek Radio Observatory, PO Box 443, Krugersdorp, 1740, South Africa. [8]Department of Physics and Astronomy, York University, Toronto, Ontario M3J 1P3, Canada [9]Leiden Observatory, Leiden University, Postbus 9513, 2300 RA Leiden, The Netherlands. [10]Netherlands Institute for Radio Astronomy (ASTRON), Postbus 2, 7990 AA Dwingeloo, The Netherlands. [11]Centre for Astrophysics and Supercomputing, Swinburne University of Technology, Hawthorn, Victoria 3122, Australia. [12]Curtin Institute of Radio Astronomy, Curtin University of Technology, GPO Box U1987, Perth, Western Australia 6845, Australia.

‡Deceased.

**The class of type Ic supernovae have drawn increasing attention since 1998 owing to their sparse association (only four so far) with long duration γ-ray bursts (GRBs)[1–4]. Although both phenomena originate from the core collapse of a massive star, supernovae emit mostly at optical wavelengths, whereas GRBs emit mostly in soft γ-rays or hard X-rays. Though the GRB central engine generates ultra-relativistic jets, which beam the early emission into a narrow cone, no relativistic outflows have hitherto been found in type Ib/c supernovae explosions, despite theoretical expectations[5–7] and searches[8]. Here we report radio (interferometric) observations that reveal a mildly relativistic expansion in a nearby type Ic supernova, SN 2007gr. Using two observational epochs 60 days apart, we detect expansion of the source and establish a conservative lower limit for the average apparent expansion velocity of 0.6$c$. Independently, a second mildly relativistic supernova has been reported[9]. Contrary to the radio data, optical observations[10-13] of SN 2007gr indicate a typical type Ic supernova with ejecta velocities ~6,000 km s$^{-1}$, much lower than in GRB-associated supernovae. We conclude that in SN 2007gr a small fraction of the ejecta produced a low-energy mildly relativistic bipolar radio jet, while the bulk of the ejecta were slower and, as shown by optical spectro-polarimetry[14], mildly aspherical.**

On 2007 August 15.51 UT the Katzman Automatic Imaging Telescope (KAIT) discovered[15] SN 2007gr at magnitude 13.8 in the bright spiral galaxy NGC 1058, at a distance[16] of 10.6 ± 1.3 Mpc. At discovery, SN 2007gr was less than five days old, based on its non-detection with KAIT on 2007 August 10.44 UT. Later optical observations[11,12] firmly classified SN 2007gr as a type Ic stripped envelope core-collapse supernova. SN 2007gr was one of the closest of its kind. Radio observations[17] with the Very Large Array (VLA) on 2007 August 17 revealed a radio source with a flux density $F_{8.4\text{GHz}} = 0.610 \pm 0.040$ mJy, thus making the source an ideal candidate for high-resolution radio imaging. The electronic very long baseline interferometry (e-VLBI) technique, which significantly improved the flexibility of the European VLBI network (EVN), enabled us to carry out sensitive high-angular-resolution observations soon after the discovery.

We observed SN 2007gr at 5 GHz with a subset of the EVN on 2007 September 6–7 for 11 h (~25 days after the supernova explosion), using the e-VLBI technique (see also Supplementary Information Sections 1 and 3). We detected a source with a peak brightness of 422 µJy per beam at 5.6 times the off-source noise level of 75 µJy per beam and determined an upper limit of 7 milliarcseconds (mas) for its angular diameter size (Fig. 1). At 10.6 Mpc, this corresponds to a linear (diameter) size of $<1.1 \times 10^{18}$ cm, which sets an upper limit of $\langle v_{app} \rangle < 8.6c$ on the average isotropic apparent expansion speed of the ejecta.

Our second observation was carried out with the EVN and the Green Bank Telescope (GBT) on 2007 November 5–6 for 10 h, ~85 days after the explosion (see also Supplementary Information Sections 2 and 3). We detected the supernova with a peak brightness of 60 µJy per beam (4.7$\sigma$) at 5 GHz with this sensitive VLBI network (Fig. 1). The observed peak brightness was significantly below the 260 µJy total flux density measurement of the Westerbork Synthesis Radio Telescope (WSRT), which is part of the VLBI network. Further confirmation of the WSRT flux density measurement was obtained from archival VLA data (Fig. 2), taken just 13 days after the second VLBI epoch, which measured a flux density of 250 ± 40 µJy at 5 GHz. Figure 2 shows a peak flux density of ~1 mJy at ~5 days, which corresponds to a peak luminosity of $1.3 \times 10^{26}$ erg s$^{-1}$ Hz$^{-1}$. Compared to other 'normal' Ib/c radio supernovae[18], this luminosity is at the lower end of the distribution (see also ref. 9).

Although part of the discrepancy between the VLBI peak brightness and the WSRT and VLA flux densities may be attributed to phase coherence losses, data simulations showed that this cannot explain such a dramatic decrease in the source peak brightness (see also Supplementary Information Section 4). Moreover, the first epoch e-VLBI run used exactly the same observing scheme, and it did not show any discrepancy between the VLBI and WSRT measurements. The best explanation for the apparently low peak brightness is that the source was resolved by the second VLBI observation. Under this assumption, we derive a conservative lower limit for the angular diameter size of 1.7 mas, which is the geometric mean of the major and minor axes of the beam. This corresponds to a linear size of $2.7 \times 10^{17}$ cm, and a lower limit of $\langle v_{app} \rangle > 0.61c$ on the average apparent expansion speed (and $v > 0.52c$ for the true expansion speed). These results conservatively assume isotropic expansion (that is, half the diameter travelled in 85 days). Note that the actual size and average expansion speed are probably somewhat larger than this lower limit ($\langle v_{app} \rangle \approx c$ and $v \gtrsim 0.7c$). Furthermore, as we estimate the average expansion speed, and the emitting region is most likely to be decelerating at this stage (as suggested by the fading radio flux density), the initial velocity of the outflow powering this extended radio emission should be higher than this estimate, and similarly $v_{app}(85 \text{ days}) < \langle v_{app} \rangle$. Further analysis of the data exploring various resolutions hints at an extension ~2.5 mas to the northeast of the supernova position, which would correspond to a linear offset of ~$4.0 \times 10^{17}$ cm and a (one-sided) superluminal average apparent expansion velocity of ~1.8$c$.

Apart from the mildly relativistic expansion of its radio emitting material, SN 2007gr appears normal, with photospheric expansion velocities smaller[12] than those inferred in hypernovae (broad-line type Ic supernovae), some of which are associated with GRBs. The radio synchrotron emission is already fading and optically thin several days after the explosion (with $F_\nu \propto \nu^{-0.66 \pm 0.15}$ on 2007 August 17, based on the fluxes in Fig. 2 and in ref. 17; here $F_\nu$ is the spectral flux density and $\nu$ is the frequency). This observation, combined with the lower limit on the source size at ~85 days ($v_{app} \approx 0.6–1\ c$) sets a lower limit on the energy, $E_{radio}$, of the radio-synchrotron-emitting material, $E_{min} \approx (0.7–1.3) \times 10^{46}(10f)^{3/7}$ erg, where $f$ is the fractional volume within the inferred source radius filled by the relativistic electrons and magnetic fields[6]. $E_{min}$ corresponds to (almost) equipartition—equal energy in relativistic electrons and magnetic fields. A reasonable deviation from equipartition may result in $E_{radio}/E_{min} \approx 10-100$. For an initially relativistic flow of total energy $E$ occupying a fraction $f_\Omega$ of the total solid angle, and expanding into an external medium with mass density $\rho = Ar^{-2}$, where $r$ is the distance from the progenitor star, and $A$ is a normalization factor, the flow becomes non-relativistic[6] at $t_{NR} = 2.3 E_{48} A_0^{-1} f_\Omega^{-1}$ days; here $E_{48}$ is $E$ in units of $10^{48}$ erg, and $A_0$ is $A$ in units of $1.5 \times 10^{10}$ g cm$^{-1}$. Our observations indicate $t_{NR} \approx 6-27$ days, suggesting either a spherical flow ($f_\Omega = 1$) with $E_{radio} \approx 10^{49}$ erg, where the electrons and magnetic field are very far from equipartition, or more plausibly, that the mildly relativistic outflow is collimated into narrow bipolar jets (for example, $f_\Omega \approx 0.03$ and $f \approx 0.1 f_\Omega$ give $E_{radio}/E_{min} \approx 10^2$ and $E_{radio}$ as a few times $10^{47}$ erg). We note here that for SN 2007gr, $E_{min} \approx 10^{46}$ erg, which is a few times smaller than that inferred[4] for GRB 060218-SN 2006aj and a few orders of magnitude lower than the other secure GRB-SN associations[4]. Our observations thus suggest that the mildly relativistic radio-emitting material in SN 2007gr carries only a very small fraction (~$10^{-4}$) of the total explosion energy, estimated to be a few times $10^{51}$ erg from optical observations[12].

Our discovery of the peculiar radio properties of SN 2007gr has important implications for the diversity of H-stripped core-collapse type Ic supernovae. Thus far we have seen: SN 2007gr, which might be the typical case (modest energy in (mildly) relativistic ejecta and no GRB); events like SN 2006aj/GRB 060218 (very dim GRB with very modest energy in relativistic ejecta); the hypernova-like SN 1998bw/GRB 980425 (dim GRB, but slightly more energy in relativistic ejecta); and last, SN 2003dh/GRB 030329 or GRBs at redshift $z \gtrsim 1$ ($E \gtrsim 10^{51}$ erg in relativistic ejecta). It is possible that all (or at least most) type Ic supernovae produce bipolar jets that are at least mildly relativistic, but that the relativistic energy content varies dramatically while the total explosion energy is much more standard. In this picture, most type Ic supernovae are likely to have very modest energy in relativistic material (and do not produce a GRB), making their radio emission detectable (and resolvable, as in SN 2007gr) only from a small local volume that implies a low detection rate; this explains why SN 2007gr is one of the two supernovae so far (besides SN 2009bb[9]) to show evidence for mildly relativistic expansion.

**Supplementary Information** is linked to the online version of the paper at www.nature.com/nature.

**Acknowledgements** Z.P. acknowledges support from the Hungarian Scientific Research Fund (OTKA, grant K72515). We are grateful to A. Soderberg for providing the VLA coordinates of SN 2007gr in August 2007, which helped us to confirm our initial e-VLBI detection and thus enabled our follow-up VLBI observations. e-VLBI development in Europe is supported by the EC DG-INFSO funded Communication Network Developments project 'EXPReS', contract no. 02662. The EVN is a joint facility of European, Chinese, South African and other radio astronomy institutes funded by their national research councils. The WSRT is operated by ASTRON (Netherlands Institute for Radio Astronomy) with support from the Netherlands Foundation for Scientific Research (NWO). ParselTounge was developed in the context of the ALBIUS project, funded by the European Community's sixth Framework Programme under RadioNet R113CT 2003 5058187. J.G. acknowledges a Royal Society Wolfson Research Merit Award. A.J.v.d.H. was supported by the NASA Postdoctoral Program at the MSFC, administered by Oak Ridge Associated Universities through a contract with NASA.

**Author Contributions** Z.P., G.B.T. and M.B. were responsible for data analysis, C.K. for project initiation and manuscript writing, E.R.-R. and J.G. for theoretical interpretation. All other authors have significantly contributed to the data analysis and interpretation.

**Author Information** Reprints and permissions information is available at www.nature.com/reprints. The authors declare no competing financial interests. Correspondence and requests for materials should be addressed to Z.P. (zparagi@jive.nl).

**Figure 1 EVN and EVN+GBT observations of SN 2007gr.** The colours, ranging from −150 to 441 µJy per beam, show the map of SN 2007gr observed on 2007 September 6–7 at 5 GHz with the EVN using the e-VLBI technique. The off-source noise in the map is 75 µJy per beam, and the peak is 422 µJy per beam (5.6σ). The VLBI location of RA = 02 h 43 min 27.97151 s, dec. = +37° 20′ 44.6873″ (J2000) is consistent with the VLA coordinates RA = 02 h 43 min 27.972 s, dec. = +37° 20′ 44.677″ (J2000) obtained at a lower resolution. The black contours show the naturally weighted and tapered EVN+GBT image of SN2007gr on 2007 November 5–6. At this epoch the off-source noise is 13 µJy per beam, and the peak is 60 µJy per beam (4.7σ). The image is centred at the position measured by the EVN at the first epoch. The apparent position shift in the peak brightness from the centre indicates that at lower resolution there is some extended flux density detected near the supernova position. The restoring beam for this image is 15.26 × 6.85 mas at a position angle of 53.3°. The contours are drawn at −26, 26, 39 and 52 µJy per beam (corresponding to −2, 2, 3 and 4 σ).

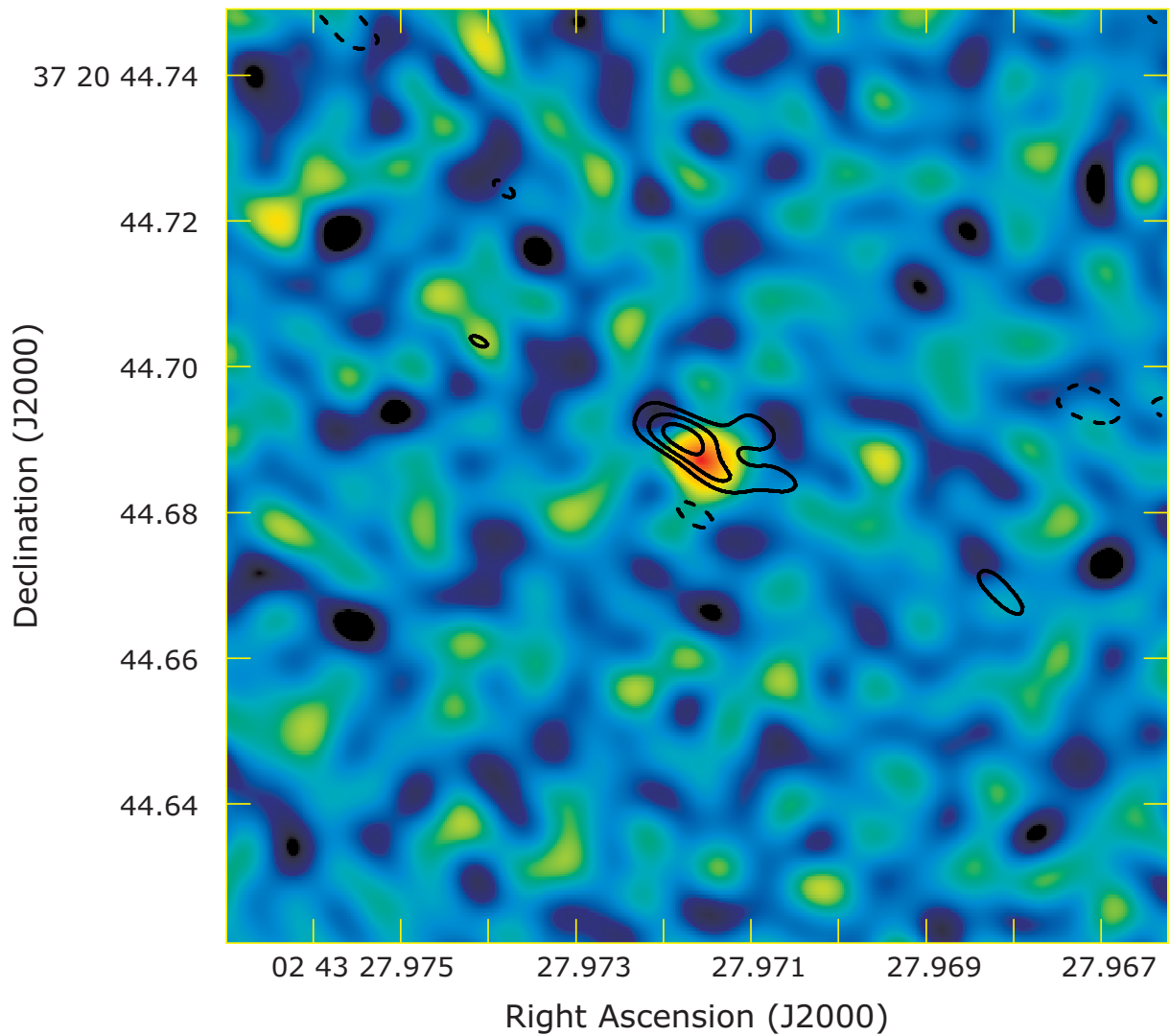

**Figure 2 VLA and WSRT light curve of SN 2007gr.** The VLA observed SN 2007gr at 4.86 GHz between 2007 August 17 and 2007 December 21 on 10 separate occasions for a typical duration of 15 min each. In the first five epochs the VLA was in A configuration, then two observations were conducted in a hybrid (BnA) configuration, and the last three observations were carried out in B configuration. Flux densities for each epoch were derived from model fitting a single Gaussian component to the visibility data. We show here the light curve including the two WSRT observations (during the VLBI observations), indicated with two vertical lines. A power-law fit to all the data except the first point when the flux density was still rising is shown (solid line) and has a slope of −0.56 ± 0.06, with a reduced $\chi^2$ value of 1.70. All times are referenced to 2007 August 13, presumed to be close to the time of the explosion. The error bars indicate $1\sigma$ uncertainties.

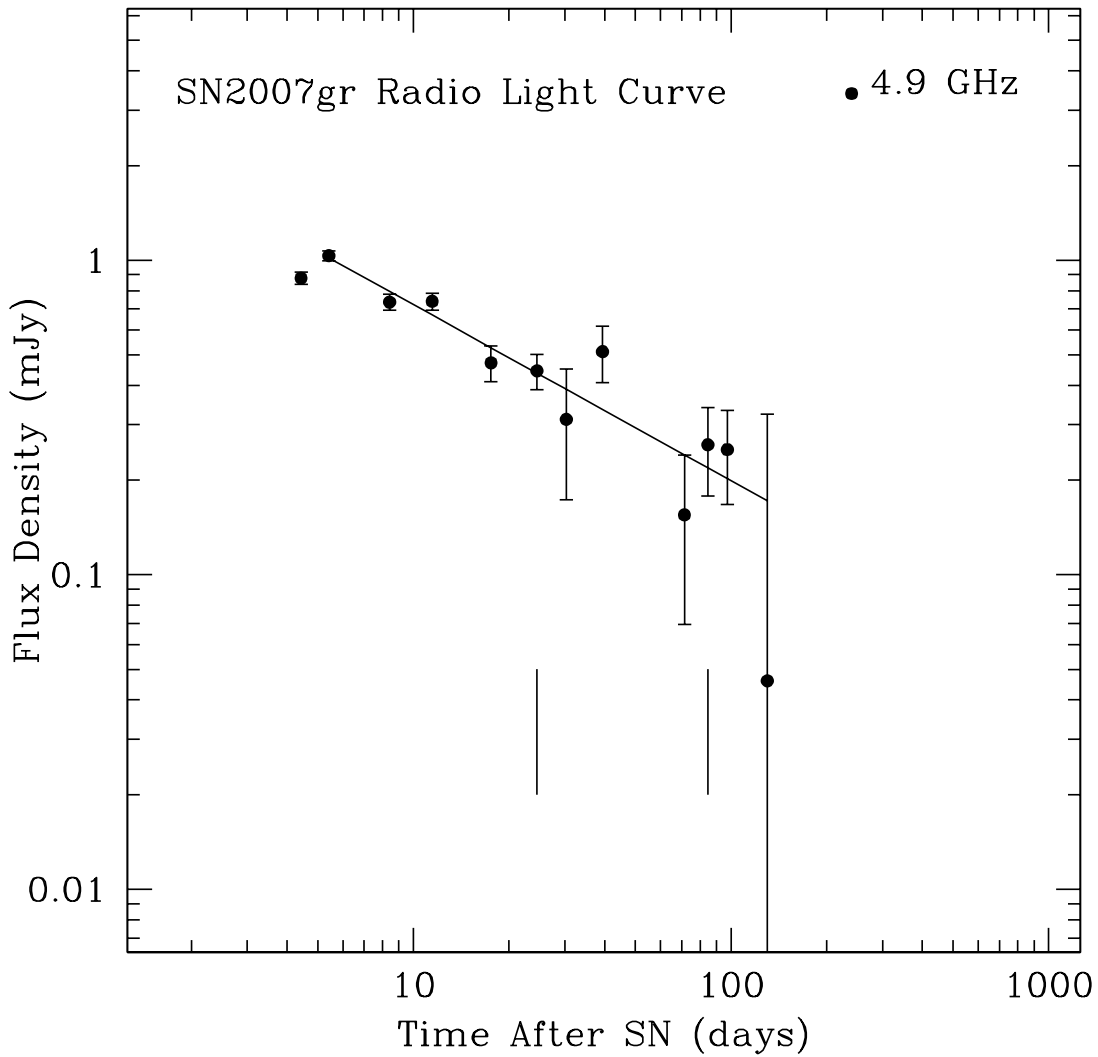

# SUPPLEMENTARY INFORMATION

## 1. First observation of SN 2007gr with the e-VLBI technique

SN 2007gr was observed at 5 GHz with a subset of the EVN on 2007 September 6-7 for 11 hours, using the e-VLBI technique. The participating telescopes were Darnhall (from MERLIN, UK), Jodrell Bank (MkII, UK), Medicina (Italy), Onsala (Sweden), Toruń (Poland) and the phased-array Westerbork Synthesis Radio Telescope (WSRT, Netherlands). The data were streamed real-time to the EVN Data Processor[19] at JIVE at 256 Mbps data rate. This was the first dedicated science observation with the 'e-EVN' in which the UDP protocol (http://www.faqs.orgrfcs/rfc768.html) was used (see also [20]), allowing us to reach 256 Mbps in spite of an unexpected network problem in the UK.

At the time of the observation some of the stations, notably the 100m Effelsberg telescope as well as telescopes in China and South Africa, did not have fast Internet connections yet, limiting somewhat the sensitivity and especially the resolution of the array compared to the full EVN. The correlation averaging time was 2 s and we used 32 delay steps (lags) per sub-band. The optical discovery coordinates were used for correlation. The positional uncertainty (order of 0.1″) was well below the achieved field of view, which was actually limited by the phased-up WSRT (~1″) rather than the correlation parameters. There were 4×8 MHz sub-bands observed in both LCP and RCP polarizations with 2-bit sampling.

The initial clock searching was carried out before the experiment using the fringe-finder source 4C39.25. SN2007gr was then phase-referenced[21] to the nearby calibrator J0253+3835 in 1.5−5 minute cycles. The J2000 coordinates of the reference source RA=$02^h$ $53^m$ $08^s.888097$, DEC=+38° 35′ 24″.99792 were taken from the Very Long Baseline Array (VLBA) Calibrator Survey (http://www.vlba.nrao.edu/astro/calib/index.shtml) which had a positional uncertainty of 0.55 mas. Additional nearby calibrators J0230+4032 and J0247+3254 as well as a primary amplitude calibrator 3C138 were observed, to calibrate the synthesis array data from the WSRT that were recorded parallel to the e-VLBI run. Short scans were scheduled every few hours on the bright fringe-finders 3C454.3 and 3C84 to ensure that the array was working as expected. In fact fringes could be monitored all along the experiment using the moderately bright (2−400 mJy) calibrators, thanks to the integrated fringe display developed by EXPReS[20], which shows the cross-correlation amplitude and phase average in the lag domain over a number of integrations.

## 2. Second SN 2007gr observation with EVN+GBT

Based on the success of the first e-VLBI run we organised a Target of Opportunity (ToO) follow-up observation with the EVN+GBT on 2007 November 5−6. The EVN observed at 1 Gbps data rate for 10 hours with Darnhall (MERLIN, UK), Jodrell Bank (MkII, UK), Effelsberg (Germany), Green Bank (US), Hartebeesthoek (South Africa), Medicina and Noto (Italy), Onsala (Sweden), Toruń (Poland) and the WSRT (Netherlands) at 5 GHz. The data were 2-bit sampled for the EVN; at the GBT we achieved an equivalent total bandwidth of 128 MHz per polarization with 512 Mbps data rate and 1-bit sampling. We followed a similar observing strategy as described above. The data in this case were recorded with Mark5As[22] and correlated at a later time at JIVE in two passes. There were 128 delay steps per sub-band and the integration time was 0.5 second. Onsala could not participate in the observation because of severe weather conditions; Toruń data were later discarded because of a technical problem.

## 3. Data analysis

Post-processing was done with the US National Radio Astronomy Observatory (NRAO) AIPS package using the standard procedures[23]. The amplitudes were calibrated using the known antenna gain curves and the measured system temperatures. The fringe-finder and phase-reference sources were fringe-fitted, and the solutions were interpolated to the target. The data were then band pass calibrated, and averaged in each sub-band. We imaged the phase-reference source with Difmap[24].

Using this model of the reference source, we corrected for its structural phase in AIPS. Imaging of the target did not involve any self-calibration. Besides this standard procedure, we further improved the amplitude calibration of the VLBI array as follows. Westerbork synthesis array data for the three phase calibrators were then analysed and the flux density scale was calibrated with 3C138. Then, by comparing the VLBI and the WSRT fluxes of the calibrators the overall amplitude scale of the VLBI data could be refined. Because even the most compact sources may have faint, extended emission resolved out on mas scales, it is desirable to use more than one calibrator. In our case, J0253+3835 had significantly larger (>10%) flux density on short spacings (WSRT) compared to VLBI, while J0247+3254 and J0230+4032 WSRT and VLBI flux densities were consistent to within a few percent after the final calibration. Should these two sources possess about the same fraction of extended flux density, this would result in an overestimation of the VLBI flux density by the same fraction. However, since using the initial gain and system temperatures the calibrator amplitudes were already in good agreement with the WSRT data (the VLBI amplitudes were initially ~10% high), we believe this effect to be significantly smaller than 10%.

In the e-EVN data on 2007 September 5-6, SN2007gr was detected with a peak brightness of 422 μJy/beam at RA=$02^h 43^m 27^s.97151$, DEC=$+37° 20' 44''.6873$ (J2000). This is 5.6 times the 1σ off-source noise level of 75 μJy/beam. The offset from the optical position is only 102 mas. The noise in the map did not reach the 5σ level elsewhere within 1″ of the optical position. The agreement with the VLA position of RA=$02^h 43^m 27^s.9725$, DEC=$+37° 20' 44''.702$ (A. Soderberg, private communication) is excellent. An estimate of the VLBI position errors is as follows: ±0.55 mas systematic (accuracy of the phase-reference source position) and ±0.61 mas statistical (beam size divided by 2×Signal over Noise Ratio), resulting in a total uncertainty of ±0.82 mas. The geometric mean of the axes of the slightly elongated beam is 6.85 mas when the data were weighted inversely proportional by the square of the noise (hereafter referred to as natural weighting). These coordinates are the most accurately determined for SN2007gr. The WSRT data show a source with a flux density of 445±40 μJy, in perfect agreement with the VLBI result.

The noise in the EVN+GBT data taken on 2007 November 5-6 was ~13 μJy/beam. The source was detected at 61 μJy/beam, which is 4.7 times the noise level, using natural weighting, with a resolution of 2.5×0.9 mas (roughly North-South orientation, beam major axis PA=−13°). The observed position is consistent with the first epoch e-VLBI position within the errors (offset 0.65 mas at PA=40°). The WSRT data show a source with a flux density of 259±40 μJy. This large discrepancy is probably not due to background emission in NGC1058. The typical mean brightness of diffuse emission measured in various boxes near the SN position is 25 μJy/beam in the WSRT maps.

## 4. Detailed investigation of the second epoch phase-referencing result

In typical EVN experiments the coherence loss may result in a 10−20% drop in target source peak brightness (e.g.[25]). Improper modelling of the array during correlation (e.g. a slight error in station position) results in the largest phase errors at the longest baselines. After removing the longest baselines (those involving the Green Bank Telescope and Hartebeesthoek) we did not obtain an increase in the peak brightness. We then investigated if un-modelled large-scale structure in the phase-reference source could have resulted in a significant phase-error, mostly affecting the short and very sensitive Westerbork-Effelsberg baseline. Fringe-fitting without using the Ef-Wb baseline did not improve our detection. To check if there were large phase-errors in the data, we performed phase-self-calibration with a point source model with a very long, 300 minutes solution interval. The coherence time after phase-referencing can be extended to at least an hour at 5 GHz[21]. With this test we identified a drift of phase in Wb and Ef during the last two hours of the experiment, when the elevation of the target was getting low in Europe. Going back to the original (not phase self-calibrated) SN 2007gr data we deleted the last two hours of visibility data, and applied uv-tapers at various radii. The target showed increasing brightness with increasing beam size up to ~100 µJy/beam, while the peak brightness location shifted up to about 2.5 mas along the position angle of PA ≈ 36˚. The tapered dirty map thus indicated extended emission towards the North-East. We note that although this result still does not represent the full flux density of the target, it strongly indicates the presence of extended structure. We are very cautious in the interpretation of this result, however, because of the low signal to noise ratio.

As a further test, we interpolated fringe-fit solutions of the phase-reference source to two nearby calibrators that were frequently observed during the run. These sources, J0230+4032 and J0247+3254, were located within 4.7 and 5.8 degrees away from J0253+3835, respectively.

Phase self-calibration resulted in a majority of the solutions within 40 degrees, with a few outliers, except for Hartebeesthoek. There is thus no sign of dramatic phase-coherence losses in the data. Also note that the target-phase-reference source separation was significantly smaller, 2.3 degrees. This means that the level of expected phase-errors in the target data is smaller. Phase self-calibration of the target data allowing for errors up to this level showed that the data are still not consistent with a single point source at the 260 µJy level.

We fitted circular Gaussian models directly to the VLBI visibility data for the second epoch. We found that an extended Gaussian with FHWM = 7 mas provides a good fit to the visibilities, and a reasonable match to the WSRT total flux density. Note however that a relatively compact model with FWHM ≥ 1 mas, with a flux density of only 160 µJy provides a somewhat better fit to the visibility data, therefore model-fitting alone - not taking into account the total flux density constraint - provides only ambiguous support to the hypothesis of a source size larger than 1.7 mas.

We also simulated visibility data from a 260 µJy point source with identical array configuration. An AIPS SN table was produced with Gaussian phase errors using ParselTounge[26]. Random phase errors with a standard deviation of 10-20 degrees, consistent with the observed errors, did not result in very significant correlation losses. Standard deviation of 45 degrees (much higher than expected) resulted in significant coherence losses, but still less than 50%. Our conclusion is that the observed data are not consistent with a single 260 µJy point source, and that a significant part of the flux density is in extended structure. The map shown in Fig. 1 is a possible representation of that structure, although the details (cleaned flux density, peak brightness, position) are admittedly affected by modest phase errors in the data.

19. Schilizzi, R. T., *et al,* The EVN-MarkIV VLBI Data Processor. *Experimental Astronomy,* **12**, 49−67 (2001).